\documentclass{ws-procs975x65}

\begin{document}

\title{VACUUM POLARISATION ON THE BRANE\\ FOR A HIGHER DIMENSIONAL\\ BLACK HOLE SPACETIME}

\author{MATTHEW HEWITT$^*$ and ELIZABETH WINSTANLEY}

\address{School of Mathematics and Statistics,The University of Sheffield,\\
Hicks Building, Sheffield, S3 7RH, UK\\
$^*$E-mail: app10mh@sheffield.ac.uk\\
www.shef.ac.uk}

\begin{abstract}
We present the vacuum polarisation of a massless, conformally coupled scalar field on the brane for a Schwarzschild--Tangherlini black hole in a bulk of zero to seven additional dimensions.
\end{abstract}


\bodymatter

\section{Vacuum Polarisation near Black Holes in Higher Dimensions}
\label{sec1}
Within the field of quantum field theory in curved space a key point of interest is the renormalised stress energy tensor (RSET) for quantised matter on black holes. The first step, and major building block, in evaluating the RSET is the calculation of the vacuum polarisation (VP) of the matter field $\phi$, equivalent to the field's auto-correlation function $\langle \phi^2 \rangle$. In addition the VP is interesting in its own right due to providing information about energy density divergences. The VP has been calculated for a variety of black hole spacetimes but has received little attention in higher dimensions. We calculate the VP on the brane for a massless, conformally coupled scalar field on a Schwarzschild--Tangherlini \cite{Tangherlini63} black hole to see what, if any, difference the extra dimensions make.

\section{Brane Worlds}
\label{sec2}
We use the ADD\cite{Arkani98,Antoniadis:1998ig} brane world model in which our universe is treated as a flat, tensionless $3+1$ brane embedded in a bulk of flat, space--like extra dimensions. These additional dimensions are considered to be large enough compared to the black hole such that any periodicity may be ignored. Our black hole is centred on and is bisected by the brane.  On the brane the four-dimensional metric reads
\begin{equation}
\label{metric}
ds^2\,=\, -f(r)dt^2+f(r)^{-1}dr^2+r^2(d\theta^2+\sin^2{\theta}\,d\phi^2) ,\qquad
f(r)\equiv 1-\left(\frac{r_h}{r} \right)^{d-3},
\end{equation}
where $r_{h}$ is the horizon radius.
 Such a brane world, though simple, is applicable to modelling the production of micro--black holes\cite{Park12}.
 We consider a massless, conformally coupled scalar field on the metric (\ref{metric}), satisfying the Klein--Gordon equation
\begin{equation}
\label{KG}
\left(\square-R/6 \right)\phi=0
\end{equation}
where $R$ is the Ricci scalar.
We calculate the VP for no extra dimensions (ie. the Schwarzschild metric, to confirm agreement with previous work\cite{Candelas84}) up to an additional seven dimensions.

\section{Calculation on the Brane}
\label{sec3}
We calculate the VP for the Hartle-Hawking vacuum\cite{Hartle76} representing a thermal state periodic in imaginary time. We apply a Wick rotation to the time coordinate and now work in a Euclidean spacetime. As any naive calculation of $\langle \phi^2 \rangle$ will be divergent, we employ point splitting\cite{Anderson90} and use a Hadamard expansion for calculating the divergent terms\cite{Decanini08}. Subtracting the renormalisation terms from the Green's functions and bringing the points together gives us the renormalised VP. This may be split into two parts: the first can be handled analytically, the second requires numerical methods. The details of the numerical work can be found in Ref.~\refcite{prep}.

\section{Results}
\label{sec4}
The calculation for zero extra dimensions has been performed previously\cite{Candelas84} and showed that the total VP differed from its analytic component by $0.69\%$ allowing the more manageable analytic terms to be a good approximation to the whole. Figure~\ref{graphs} compares the analytic and total results of the VP for a bulk total of four to eleven dimensions. These results clearly demonstrate that the analytic approximation to the whole rapidly breaks down and worsens as dimension increases, with a $52\%$ difference for eleven total dimensions.

\def\figsubcap#1{\par\noindent\centering\footnotesize(#1)}
\begin{figure}
\begin{center}
  \parbox{2.1in}{\epsfig{figure=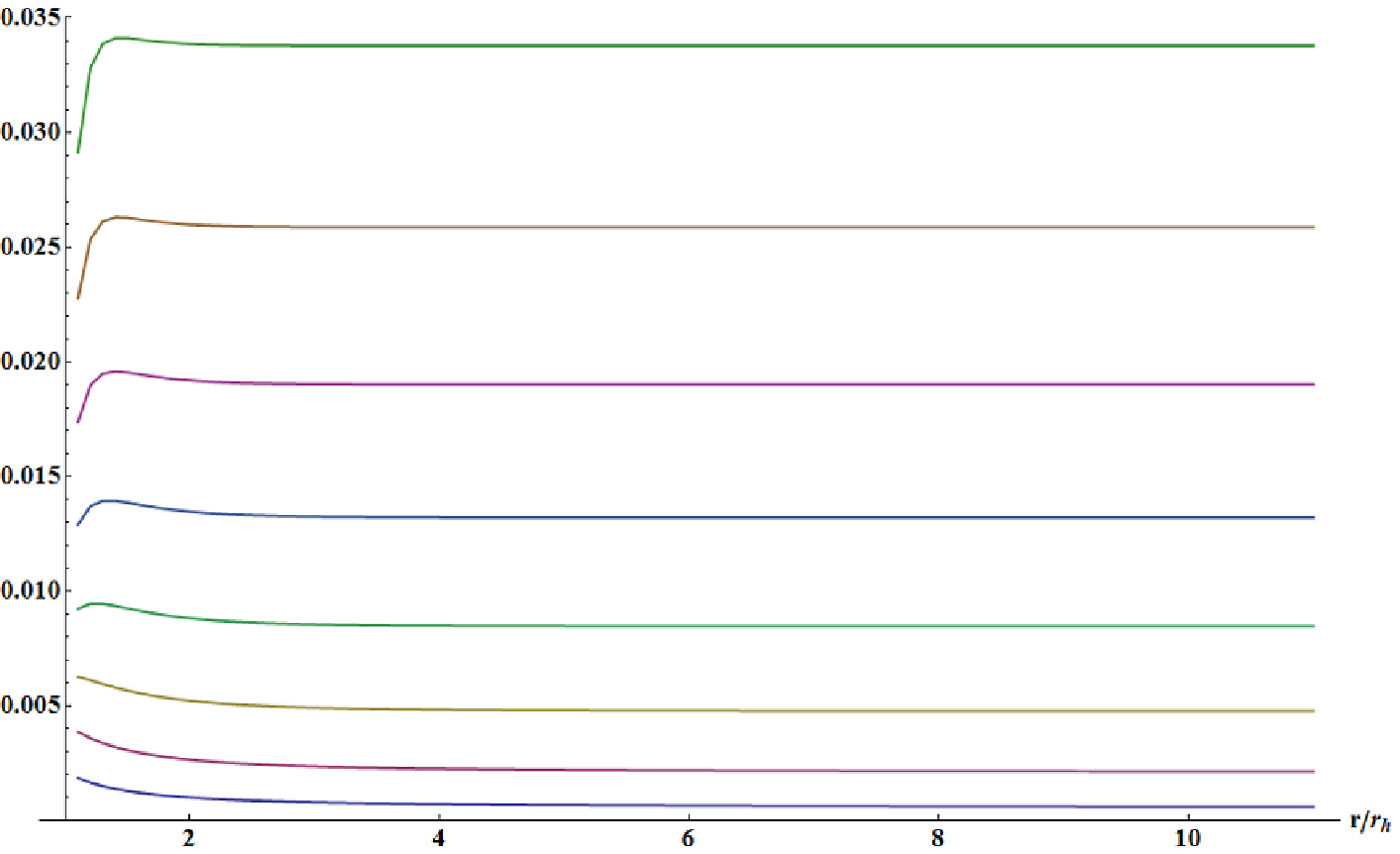,width=2in}\figsubcap{a}}
  \hspace*{4pt}
  \parbox{2.1in}{\epsfig{figure=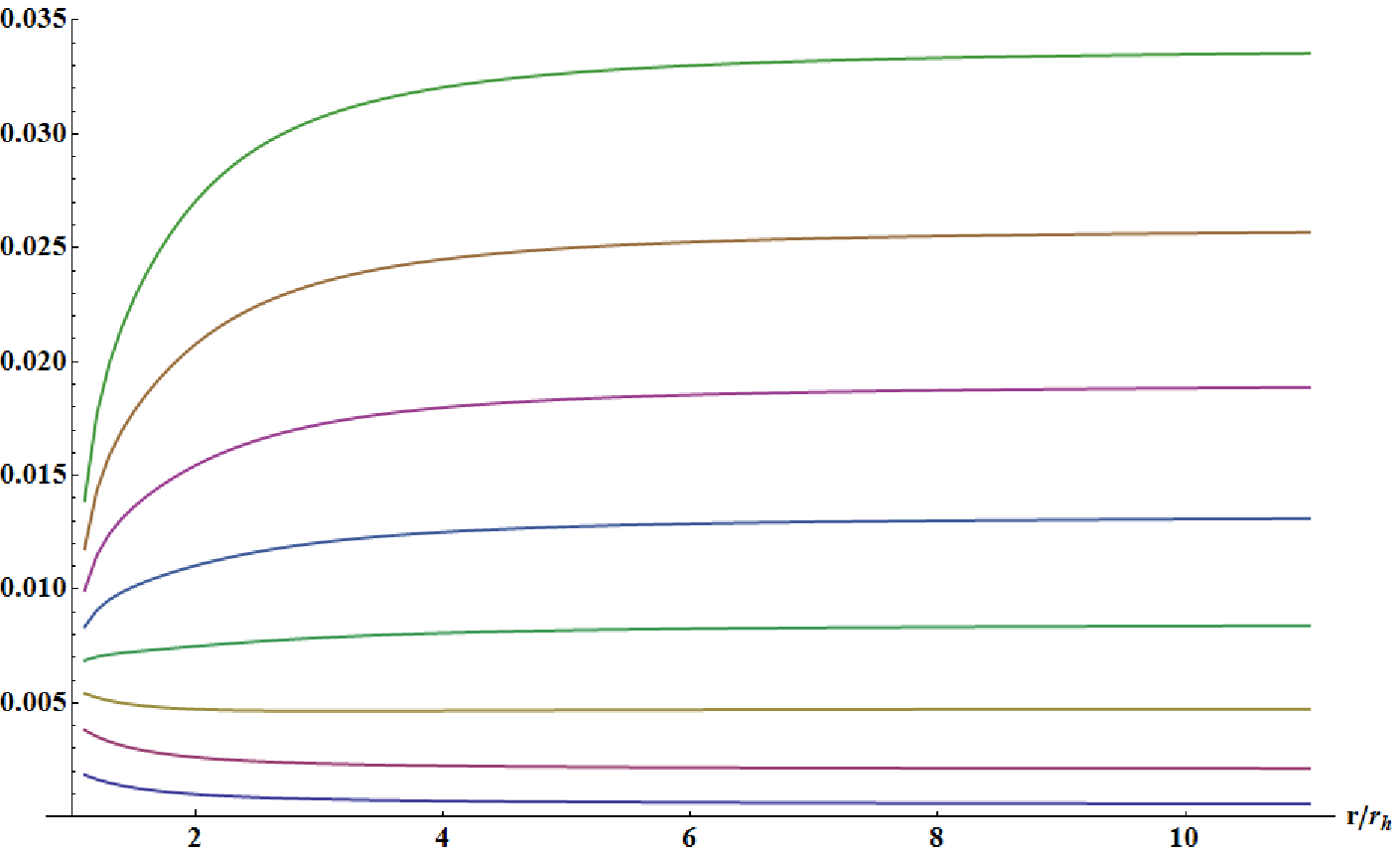,width=2in}\figsubcap{b}}
  \caption{VP plotted as a function of $r/r_h$ such that the horizon is the far left vertical. Both plots are on the same scale. The number of extra dimensions increases up the plots such that on both the lowest line is for zero extra dimensions and the top seven extra dimensions. (a) VP for the analytic component only. (b) Total calculated VP.}%
  \label{graphs}
\end{center}
\end{figure}

\def\figsubcap#1{\par\noindent\centering\footnotesize(#1)}
\begin{figure}
\begin{center}
  \parbox{2.1in}{\epsfig{figure=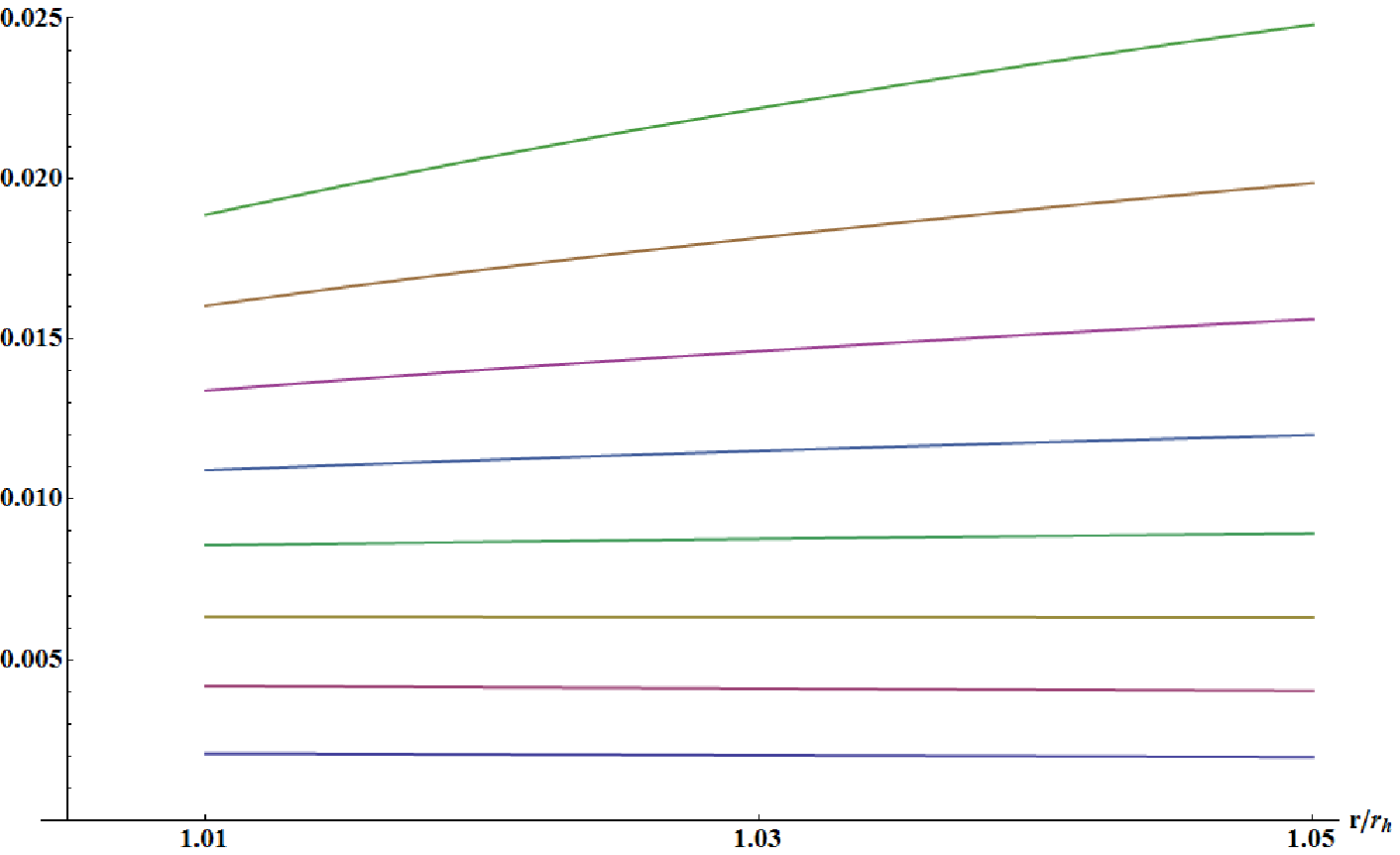,width=2in}\figsubcap{a}}
  \hspace*{4pt}
  \parbox{2.1in}{\epsfig{figure=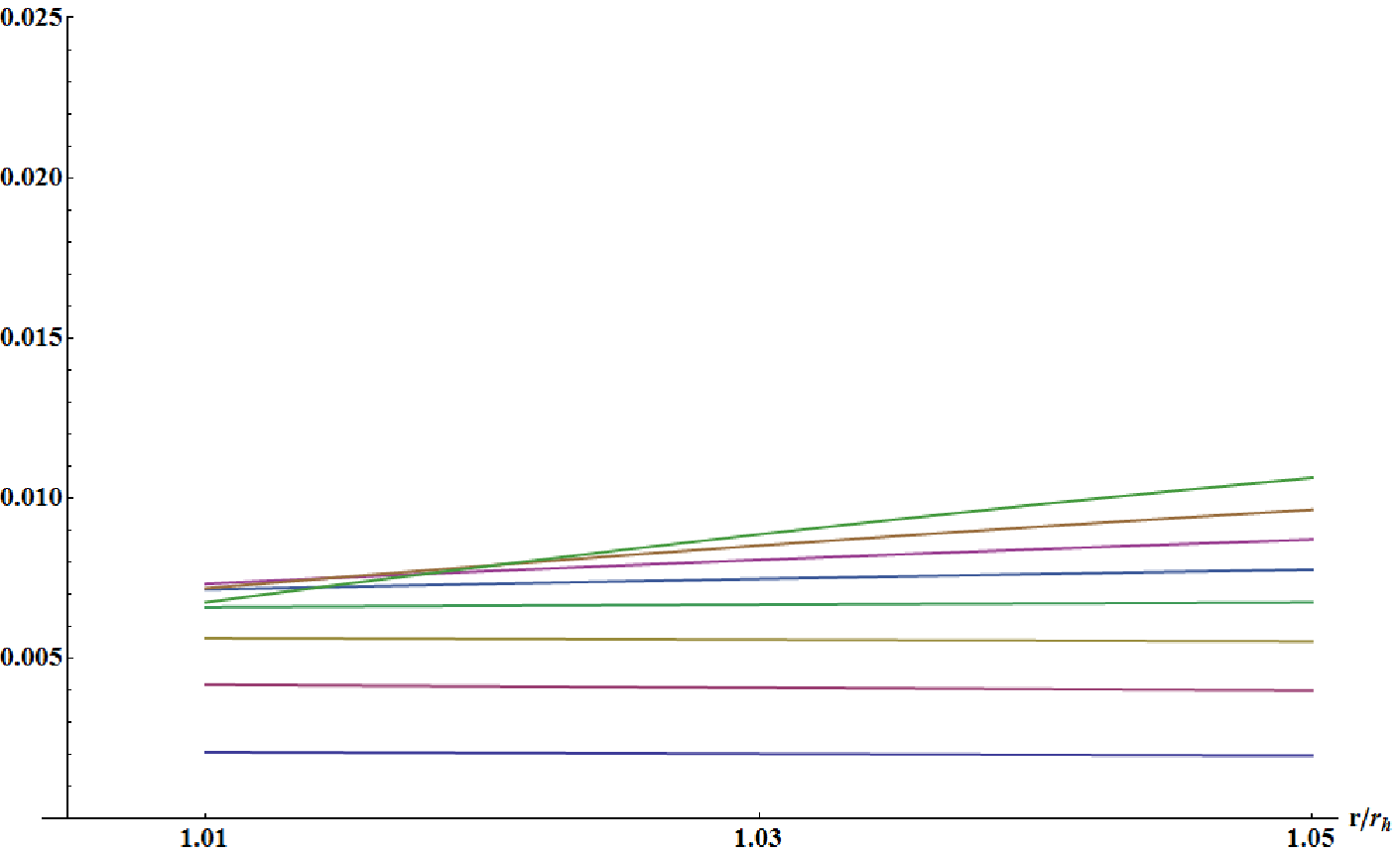,width=2in}\figsubcap{b}}
  \caption{VP near the horizon plotted as a function of $r/r_h$ such that the horizon is the far left vertical. Both plots are on the same scale. The number of extra dimensions increases up the plots such that on both the lowest line is for zero extra dimensions and the top seven extra dimensions. (a) VP for the analytic component only. (b) Total calculated VP.}%
  \label{on}
\end{center}
\end{figure}

As the major differences in value are close to the event horizon we performed the calculations again focusing on the region where $1<r/r_h<1.5$. The results of this second calculation are displayed in \fref{on}, again comparing the analytic terms against the total VP.
Figure~\ref{on} shows that the total VP falls increasingly quickly with additional dimensions. This strongly indicates that the numeric component causes the total VP to achieve a maximum on the horizon between a total of eight or nine dimensions before rapidly falling towards negative values.

\section{Further Investigation}
\label{sec5}
Having completed this calculation on the brane our methods must next be extended to calculation of the VP in the bulk. The greatest challenges in the bulk will be extending our renormalisation terms into higher dimensions (this was not required on the brane) and finding an appropriate form of our Green's function in which to incorporate them. Once the VP has been calculated in the brane and the bulk a potential extension to the work is to confirm the exact VP values on the event horizon itself but we will return this question in the future.

\section*{Acknowledgements}
This work is supported by a studentship from EPSRC, the Lancaster--Manchester--Sheffield Consortium for Fundamental Physics under STFC grant ST/J$000418/1$ and European Cooperation in Science and Technology (COST) action MP$0905$ ``Black Holes in a Violent Universe''.

\bibliographystyle{ws-procs975x65}
\bibliography{Writebib}

\end{document}